\documentclass[preprint,showpacs,pra,superscriptaddress]{revtex4-1}
\usepackage[latin9]{inputenc}
\usepackage{amsmath}
\usepackage{amssymb}
\usepackage{graphicx}

\makeatletter

\providecommand{\tabularnewline}{\\}

 \@ifundefined{textcolor}{}
 {%
   \definecolor{BLACK}{gray}{0}
   \definecolor{WHITE}{gray}{1}
   \definecolor{RED}{rgb}{1,0,0}
   \definecolor{GREEN}{rgb}{0,1,0}
   \definecolor{BLUE}{rgb}{0,0,1}
   \definecolor{CYAN}{cmyk}{1,0,0,0}
   \definecolor{MAGENTA}{cmyk}{0,1,0,0}
   \definecolor{YELLOW}{cmyk}{0,0,1,0}
 }

\makeatother

\begin{document}

%
\title{Measurement-device-independent quantum key distribution with ensemble-based memories}


\author{Nicol\'o Lo Piparo}

\affiliation{School of Electronic and Electrical Engineering, University of Leeds,
Leeds, UK}
\author{Mohsen Razavi}

\affiliation{School of Electronic and Electrical Engineering, University of Leeds,
Leeds, UK}
\author{Christiana Panayi}

\affiliation{School of Electronic and Electrical Engineering, University of Leeds,
Leeds, UK}




\begin{abstract}
Quantum memories are enabling devices for extending the reach of quantum key distribution (QKD) systems. The required specifications for memories are, however, often considered too demanding for available technologies. One can change this mindset by introducing memory-assisted measurement-device-independent QKD (MDI-QKD), which imposes less stringent conditions on the memory modules. It has been shown that, in the case of {\em fast} single-qubit memories, we can reach rates and distances not attainable by single no-memory QKD links. Single-qubit memories, such as single atoms or ions, have, currently, too slow of an access time to offer an advantage in practice. Here, we relax that assumption, and consider ensemble-based memories, which satisfy the main two requirements of having short access times and large storage-bandwidth products. Our results, however, suggest that the multiple-excitation effects in such memories can be so detrimental that they may wash away the scaling improvement offered by memory-equipped systems. We then propose an alternative setup that can in principle remedy the above problem. As a prelude to our main problem, we also obtain secret key generation rates for MDI-QKD systems that rely on imperfect single-photon sources with nonzero probabilities of emitting two photons. 
\end{abstract}

\maketitle


\section{Introduction}
%
%
%
%
Future quantum communication networks may well rely on quantum repeater links for distributing entanglement between different nodes. Such entangled states can then be used for various applications including quantum key distribution (QKD). While progress toward building repeater systems is underway, one can think of intermediary steps that can be implemented in a nearer future. On the one hand, they ease the way for future generations of quantum networks \cite{Munro:NatPhot:2012,Liang:NoMemRep_PRL2014}, and, on the other, they offer services over a range of distances not currently available by conventional direct QKD links. Memory-assisted measurement-device-independent QKD (MDI-QKD) has recently been proposed with the above objectives in mind \cite{Brus:MDIQKD-QM_2013, Panayi_NJP2014}. Such systems will resemble a single-node quantum repeater link with quantum memories (QMs) in the middle node. There is, however, no QMs at the users' ends and they are only equipped with encoder/source modules. Instead of distributing entanglement over elementary links, users send BB84-encoded states toward the memories, and once both memories are loaded with relevant states, an entanglement swapping operation is performed on the memories. In a recent work \cite{Panayi_NJP2014}, it has been shown that if one uses fast memories with large storage-bandwidth products, it would be possible to beat existing no-memory QKD systems in a practical range of interest using memories mostly attainable with current technologies. Among different developing technologies for QMs, ensemble-based memories have a good chance to satisfy both required conditions. Writing times as short as 300~ps and  bandwidths on the order of GHz have been reported for such memories \cite{Walmsley:PRL:2010, Tittel:AFCmem_Nature2011}. They are however inflicted by multiple-excitation effects, which may cause errors in QKD setups relying on such QMs. Here, we show how sensitive the performance of memory-assisted MDI-QKD can be to this type of errors and propose a modified setup resilient to multiple-excitation effects.

\begin{figure}[htbp]
  \centering
  \includegraphics[width=\linewidth]{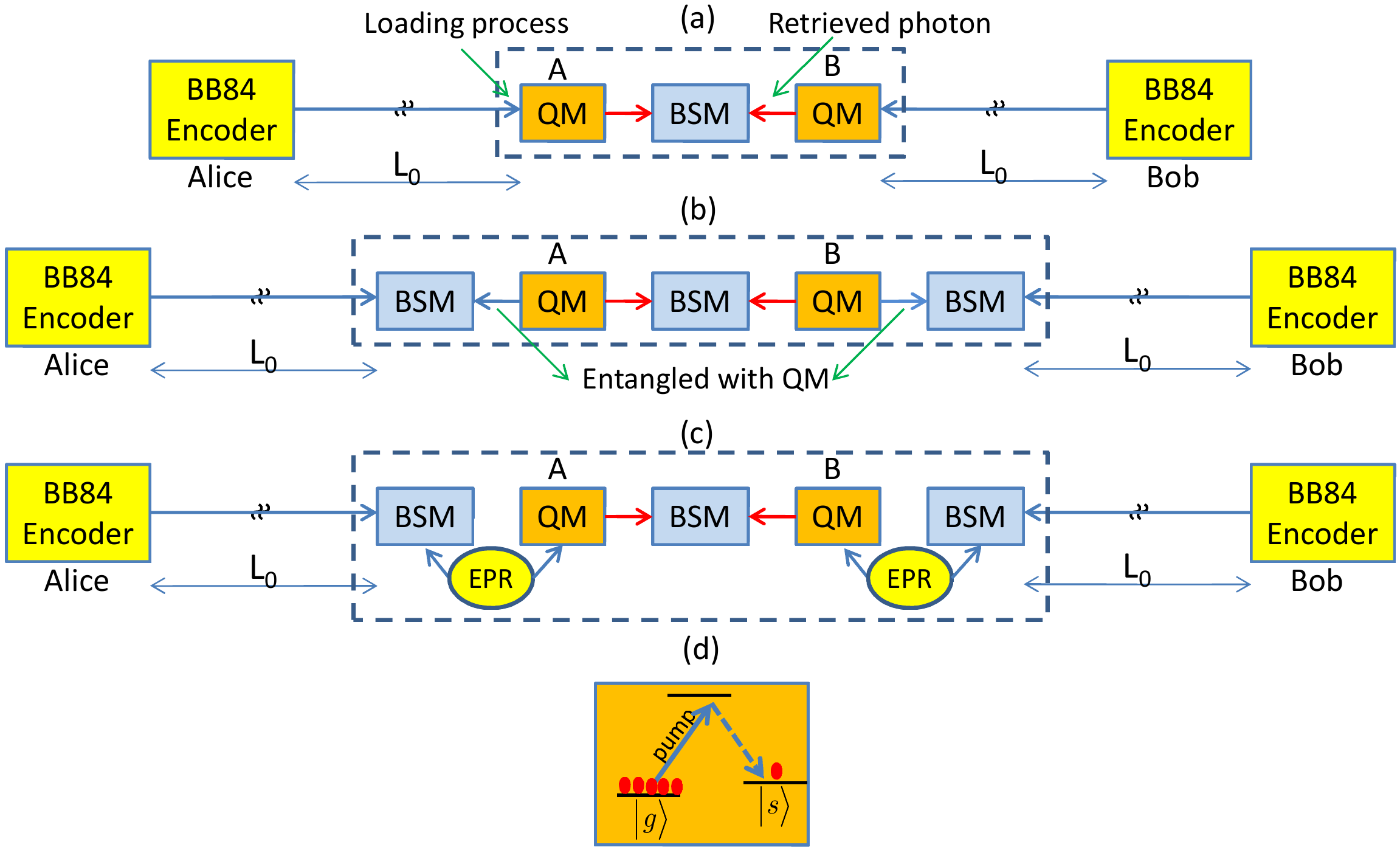}
\caption{Different setups for memory-assisted MDI-QKD. (a) MDI-QKD with directly heralding quantum memories~\cite{Panayi_NJP2014}. (b) MDI-QKD with indirectly heralding quantum memories ~\cite{Panayi_NJP2014}. At each round, an entangling process is applied to each QM, generating a photon entangled with the QM. These photons interfere at the side BSM modules next to the QMs with incoming pulses from the encoders. (c) Similar to (b), but the entanglement between the QM and a photon is achieved by generating a pair of entangled photons by the EPR source, and storing one of the photons in the QM. (d) A possible energy-level configuration for an ensemble-based QM suitable for phase encoding.}
\label{Fig:setups}
\end{figure}

MDI-QKD offers a key exchange approach resilient to detector attacks \cite{Lo:MIQKD:2012}. In this system, Alice and Bob send their encoded signals to a middle station, at which a Bell-state measurement (BSM) is performed. This BSM effectively performs an entanglement swapping operation, similar to that of quantum repeaters, on the incoming photons, based on whose result Alice and Bob can infer certain correlations between their transmitted bits. Because of relying on the reverse-EPR protocol \cite{Biham:ReverseEPR:1996}, the middle party does not need to be trusted, nor does he need to perform a perfect BSM. In the memory-assisted MDI-QKD, we add two QMs before the middle BSM module; see Fig.~\ref{Fig:setups}(a). The objective is to obtain a better rate-versus-distance behavior as now the two photons sent by Alice and Bob do not need to arrive at the BSM module in the same round. This way, we expect to get the same improvement as in single-node quantum repeaters.

The required specifications for the QMs in Fig.~\ref{Fig:setups} can be milder than that of a quantum repeater \cite{Panayi_NJP2014}. In a single-node quantum repeater, with two legs of length $L_0$ and one BSM module in the middle, we have to distribute entanglement between memories in each leg before being able to perform the BSM. For single-mode memories, the entanglement distribution scheme can only be applied once every $T_0 = L_0/c$, where $c$ is the speed of light in the channel \cite{Razavi.Lutkenhaus.09}. The required coherence time for the QMs is then proportional to $T_0$ as well. In the memory-assisted MDI-QKD of Fig.~\ref{Fig:setups}(a), the repetition rate is dictated by the writing time into QMs. If, therefore, a {\em heralding} mechanism is available, and if the QMs have short access times, we can run the MDI-QKD protocol faster than that of a quantum repeater, and, correspondingly, the required coherence time could also be lower \cite{Panayi_NJP2014}. 

The required heralding mechanism, by which we can tell if the QMs have been loaded with the corresponding state to that sent by the users, can be implemented in several ways. In Fig.~\ref{Fig:setups}(a), we rely on a direct heralding mechanism in which we attempt to store the transmitted photons into the memories and non-destructively verify whether the writing procedure has been successful. This mechanism is only applicable to a limited number of QMs, such as trapped single atoms/ions, and it is often very slow \cite{Blatt:Natture:2012}. In \cite{Panayi_NJP2014}, the authors have analyzed an indirect heralding mechanism as in Fig.~\ref{Fig:setups}(b) in the single-excitation regime, that is, when QMs can only store a qubit. In this scheme, a photon is first entangled with the QM, and then immediately a side BSM is performed on this photon and the signal sent by the user. A successful side BSM, declared by two detector clicks, ideally teleports the user's state onto the QM and heralds a successful loading event. In order to outperform no-QM QKD systems, the setup of Fig.~\ref{Fig:setups}(b) must be equipped with memories with large storage-bandwidth products as well as short access and entangling times. It turns out that the state of the art for single-qubit memories, e.g., single atoms \cite{Rempe:Nature:2012} or ions \cite{Blatt:Natture:2012}, is not yet sufficiently advanced to meet the requirements of practical memory-assisted protocols. In particular, we need faster memories for the practical ranges of interest.

Here we extend the analysis in \cite{Panayi_NJP2014} to the case of {\em ensemble-based} memories, which often offer very large bandwidths, or, equivalently, very short access times, suitable for the memory-assisted scheme. Such memories, however, suffer from multiple-excitation effects, which we carefully look into in this paper. In fact, when multiple-excitations are present, a seemingly successful side BSM may have been resulted from two photons originating from the QM in Fig.~\ref{Fig:setups}(b), in which case the final measurement results have no correlation with the transmitted signal by the user. Our results show that such effects can be so detrimental that we cannot beat no-memory QKD systems within practical ranges of interest.  We then look at an alternative indirect heralding mechanism, see Fig.~\ref{Fig:setups}(c), and show that, in principle, we can avoid multiple-excitation errors if a proper entangled-photon (EPR) source is used \cite{Berlin14}.

The rest of this paper is organized as follows. As the first step toward the analysis of the MDI-QKD system of Fig.~\ref{Fig:setups}(b) with non-qubit memories, in Sec.~\ref{Sec:ImpSources}, we study a no-QM MDI-QKD link that uses imperfect sources, that is, the ones which have a nonzero probability for generating more than one photon. This is a good approximation to the state of the field entangled with an ensemble-based QM. We then extend our results, in Sec.~III, to the memory-assisted system in Fig.~\ref{Fig:setups}(b) and study the system performance in the presence of multiple excitations in the QMs. We then propose a modified setup that can handle multiple-excitation errors. We conclude the paper in Sec.~IV commenting on the practicality of each scheme.

\section{MDI-QKD with Imperfect Sources}
\label{Sec:ImpSources}
Regardless of the type of material used, an ensemble-based memory can be modeled as a non-interacting ensemble of quantum systems. Here, for simplicity, but without loss of generality, we assume our QM is an ensemble of neutral atoms with the $\Lambda$-level configuration shown in Fig.~\ref{Fig:setups}(d). One possible way to entangle a photon with such a QM is to pump all the atoms in the ensemble to be initially in their ground states $|g\rangle$; we then excite the ensemble by a short pulse in such a way that the probability, $p$, of driving an off-resonant Raman transition in the ensemble is kept well below one. In that case, the joint state of the released Raman optical field and the ensemble follows that of a two-mode squeezed state given by \cite{Razavi.DLCZ.06}
\begin{equation}
\label{init_joint}
|\psi\rangle_{AP} = \sum_{n=0}^{\rm \# atoms}{\sqrt{(1-p)p^n}|n\rangle_A|n\rangle_P},
\end{equation}
where $|n\rangle_P$ is the Fock state for $n$ photons and $|n\rangle_A$ is the symmetric collective state to have $n$ atoms in their $|s\rangle$ states; see Fig.~\ref{Fig:setups}(d). Assuming $p \ll 1$, we can truncate the above state at $n=2$ without losing much accuracy. Furthermore, assuming that there is a post-selection mechanism by which the state $|0\rangle_A |0\rangle_P$ is selected out, the effective state for the photonic system $P$ is given by
\begin{equation}
\label{inp_st}
\rho_P(p) = (1-p)|1\rangle_P {}_P\langle 1| + p |2\rangle_P {}_P\langle 2|,
\end{equation}
which resembles an imperfect single-photon source with a nonzero probability $p$ for emitting two photons. This is the type of state that one would get for the photons entangled with the QMs in Fig.~\ref{Fig:setups}(b). That is, each leg of the system, can be modeled as an asymmetric MDI-QKD link, where the source on one side generates photons in the form of \eqref{inp_st}. The source on the user's end could be the same, or one may use decoy coherent states for practical purposes. The latter case will be investigated in a separate publication \cite{Nicolo_paper3}. Note that the type of states as in \eqref{init_joint} do not represent maximally entangled states. One can, however, combine two such states and obtain an effective entangled states after post-selection \cite{DLCZ_01}. 

In this section, we study an MDI-QKD link with imperfect sources as in \eqref{inp_st}. Although we digress a bit from the main problem, it gives us some insight into the analysis of the setup in Fig.~\ref{Fig:setups}(b), and, more generally, when MDI-QKD links are connected to quantum repeater setups \cite{Nicolo_paper3}. The type of memory considered here best fits into phase-encoded QKD setups as we will consider next \cite{MXF:MIQKD:2012}.

\subsection{Phase-encoded MDI-QKD}
\label{Sec:PhaseEnc}
\begin{figure}
\begin{centering}
\includegraphics[width=8.6cm]{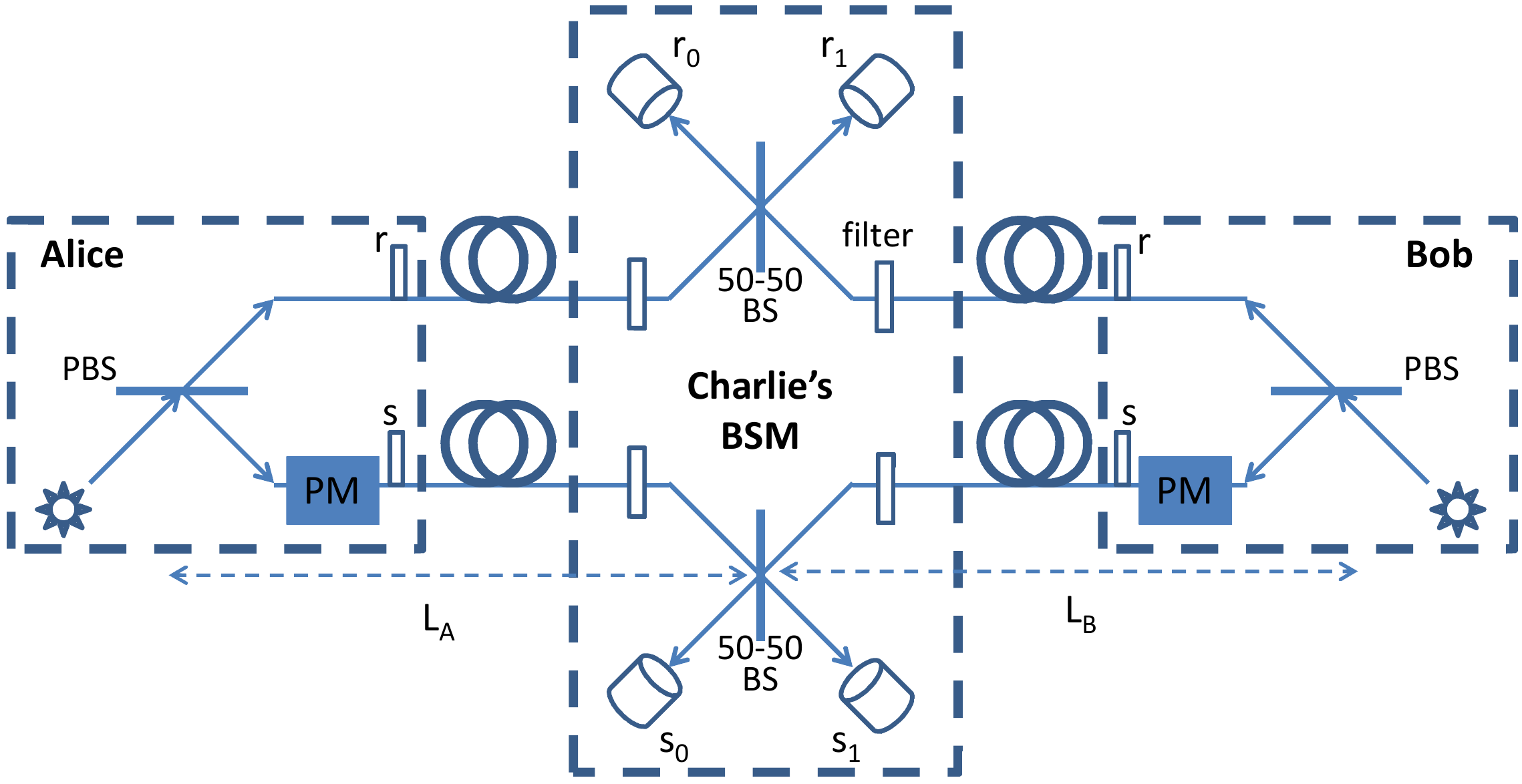}
\par\end{centering}
\caption{{\footnotesize \label{fig:Diagram-for-MDI-QKD} Schematic diagram for the MDI-QKD
protocol with phase encoding~\cite{MXF:MIQKD:2012}. Here BS stands for beam splitter, PBS for polarizing BS, and PM for phase modulator. }}
\end{figure}

In this section we describe phase-encoded MDI-QKD as proposed
in \cite{MXF:MIQKD:2012}. For the sake of convenience, we analyze the dual-rail setup in Fig. \ref{fig:Diagram-for-MDI-QKD}, but, for practical purposes, it is possible to implement the same
scheme via time multiplexing, by using only one physical channel \cite{MXF:MIQKD:2012}.  
Here, states sent by Alice and Bob are encoded either in the $z$ or the $x$ basis. Encoding the states in the $z$
basis is achieved by sending horizontally or vertically polarized
pulses to a polarizing beam splitter (PBS) to, respectively, generate a signal
in the $r$ or in the $s$ mode (corresponding to bits 0 or 1) in Fig.~\ref{fig:Diagram-for-MDI-QKD}.
To implement the $x$-basis encoding, +45-polarized pulses
are prepared at the source and two relative phases, $\left\{ 0,\,\pi\right\} $ corresponding to bits $\{0,1\}$, are used at the phase modulator. In this case, the PBS splits the signal
into $r$ and $s$ modes, and photons will be in a superposition of these modes. 

The procedure to establish a secret key is as follows. Alice and Bob,
who are separated by a distance $L = L_A + L_B$, choose randomly a basis from
$\left\{ x,\, z\right\} $ and a bit from $\left\{ 0,\,1\right\} $
and send a pulse to a middle site, where a BSM is performed by an untrusted
party, Charlie. We make photons indistinguishable through the filters
represented by empty boxes in Fig.~\ref{fig:Diagram-for-MDI-QKD}.
A click in exactly
one of the $r$ detectors, in Fig.~\ref{fig:Diagram-for-MDI-QKD}, and exactly one
of the $s$ detectors will correspond to a successful event. When the users both choose the $z$ basis, a successful event corresponds to complementary bits on the two ends. When
they both choose the $x$ basis, instead, a different bit assignment will follow. If they pick the same phase
then the state will be correlated and $r_{0}$ and $s_{0}$ or $r_{1}$
and $s_{1}$ will ideally click. We will refer to this detection event as type I. If they pick different phase values then
the state will be anti-correlated and $r_{0}$ and $s_{1}$ or $r_{1}$
and $s_{0}$ will ideally click. The latter pattern of clicks is referred to as type II. In either case, Charlie announces her BSM
results to Alice and Bob. Alice and Bob will compare the bases used
for all transmissions. They keep the results if they have chosen the
same basis and discard the rest. 


\subsection{Key rate analysis}

In this section, the secret key generation rate for the MDI-QKD scheme
of Fig.~\ref{fig:Diagram-for-MDI-QKD} is calculated. Here, we assume that Alice and
Bob each have an imperfect single-photon source that can emit two photons with probability $p \ll 1$ as in \eqref{inp_st};
hence, in our following analysis, we neglect $O\left(p^{2}\right)$
terms corresponding to the simultaneous emission of two photons by
both sources. We assume Alice and Bob are located at, respectively, distances $L_A$ and $L_B$ from the BSM module, and the total path loss for a channel with length $l$ is given by $\eta_{\rm ch}(l) = \exp{(-l/L_{\rm att})}$, with $L_{\mathrm{att}}=25$ km for an optical fiber channel. The secret key generation rate is then lower bounded by \cite{MXF:MIQKD:2012, MDIQKD_finite_PhysRevA2012}
\begin{equation}
R_{ss}\geq Q_{11}^{z}\left(1-h\left(e_{11}^{x}\right)\right)-Q_{pp}^{z}f\, h\left(E_{pp}^{z}\right),
\label{eq:Rss}
\end{equation}
where $Q_{11}^{z} = (1-p)^{2}Y_{11}^{z},$ with $Y_{11}^{z}$ being
the probability of a successful click pattern, in the $z$ basis, when Alice and Bob send
exactly one photon each; $e_{11}^x$ is the quantum bit error rate (QBER), in the $x$ basis, when Alice and Bob send
exactly one photon each; $Q_{pp}^z$ and $E_{pp}^z$ are, respectively, the gain and the QBER, in the $z$ basis, when Alice and Bob send the states as in \eqref{inp_st}; $f$ is the error correction inefficiency;
and, $h\left(x\right)=-x\,\log_{2}\left(x\right)-(1-x)\,\log_{2}\left(1-x\right)$
is the Shannon's binary entropy function. In \eqref{eq:Rss}, we have assumed that the efficient QKD protocol is used, in which the $z$ basis is used much more often than the $x$ basis \cite{Lo:EffBB84:2005}.

\begin{figure}
\begin{centering}
\includegraphics[width=5.6cm]{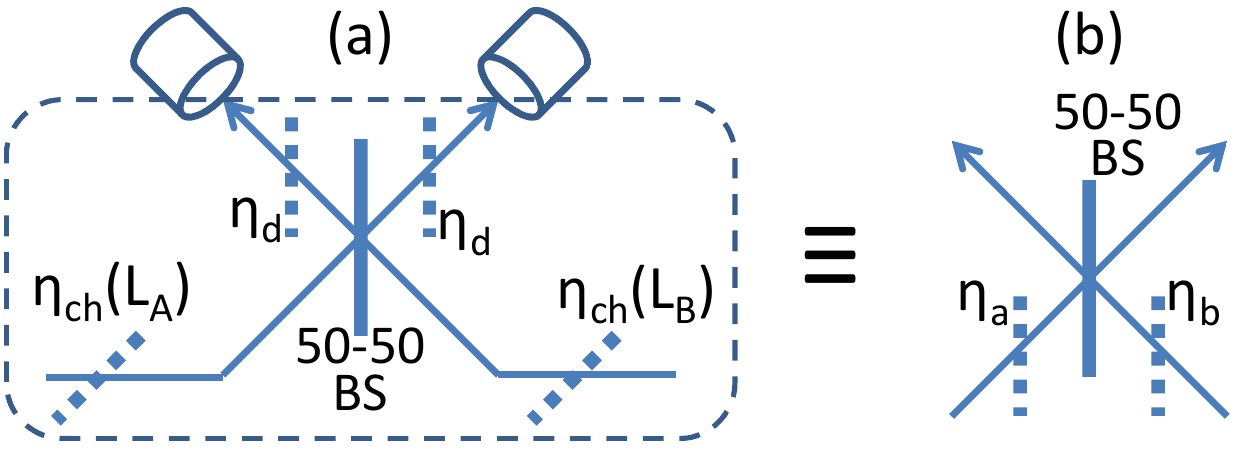}
\par\end{centering}
\caption{\label{fig:BSM}(a) The simplified module for the upper or lower arms in the setup of Fig.~\ref{fig:Diagram-for-MDI-QKD}. (b) An asymmetric butterfly module with parameters $\eta_a$ and $\eta_b$.}
\end{figure}

In Appendix A, we derive each term in \eqref{eq:Rss} under the normal mode of operation when no eavesdropper is present. This will simulate the parties' estimate of relevant parameters in the limit of infinitely long keys. We consider the dark count noise of photodetectors and possible misalignment errors in the setup. The latter will model our deviation from the indistinguishibility condition required for the BSM operation. The key tool in calculating the key rate parameters in \eqref{eq:Rss} is an asymmetric butterfly operation as shown in Fig.~\ref{fig:BSM}. By modeling the path loss in each channel as well as photodetector efficiencies, $\eta_d$, by fictitious beam splitters, each (upper or lower) arm in Fig.~\ref{fig:Diagram-for-MDI-QKD} can be modeled as in Fig.~\ref{fig:BSM}(a), in which the photodetectors have unity quantum efficiencies. This can be simplified to the butterfly module in Fig.~\ref{fig:BSM}(b), where $\eta_a = \eta_{\rm ch}(L_A) \eta_d$ and $\eta_b = \eta_{\rm ch}(L_B) \eta_d$. In Appendix A, we find the input-output relationship for all relevant input states to a general butterfly module, from which the joint state of photons sent by Alice and Bob right before photodetection can be calculated. By applying proper measurement operators on this state, we find the post-measurement state corresponding to each of the relevant click patterns. For instance, a click on the non-resolving detector $r_0$, and no click on $r_1$, can be modeled by the following measurement operator \cite{LoPiparo:2013}
\begin{equation}
\begin{array}{c}
M_{r_{0}}=(1-d_{c})\left[\left(I_{r_{0}}-|0\left\rangle _{r_{0}r_{0}}\right\langle 0|\right)\otimes|0\left\rangle _{r_{1}r_{1}}\right\langle 0|+\right.\\
\mbox{}\left.d_{c}|0\left\rangle _{r_{0}r_{0}}\right\langle 0|\otimes|0\left\rangle _{r_{1}r_{1}}\right\langle 0|\right]
\end{array},\label{eq:measurement_op}
\end{equation}
where $I_{r_{0}}$ denotes the identity operator for the mode entering
the $r_{0}$ detector, and $d_{c}$ is the dark-count rate per gate
width per detector. The measurement operator for the event that only detectors $r_0$ and $s_0$ click would then be given by $M_{r_0}\otimes M_{s_0}$, and similarly for other combinations. 

\begin{table}
\begin{centering}
\vspace{2mm}
\begin{tabular}{|c|c|}
\hline 
Quantum efficiency, $\eta_{d}$ & 0.93\tabularnewline
\hline 
Memory reading efficiency, $\eta_{r0}$ & 0.87\tabularnewline
\hline 
Dark count per pulse, $d_{c}$ & $10^{-9}$\tabularnewline
\hline 
Attenuation length, $L_{\mathrm{att}}$ & 25 km\tabularnewline
\hline 
Misalignment, $e_{d}$ & $0$\tabularnewline
\hline 
\end{tabular}
\par\end{centering}
\caption{{\footnotesize \label{tab:Nominal-values-used}Nominal values used
in our numerical results}}
\end{table}

\begin{figure}
\begin{centering}
\includegraphics[width=7.5cm]{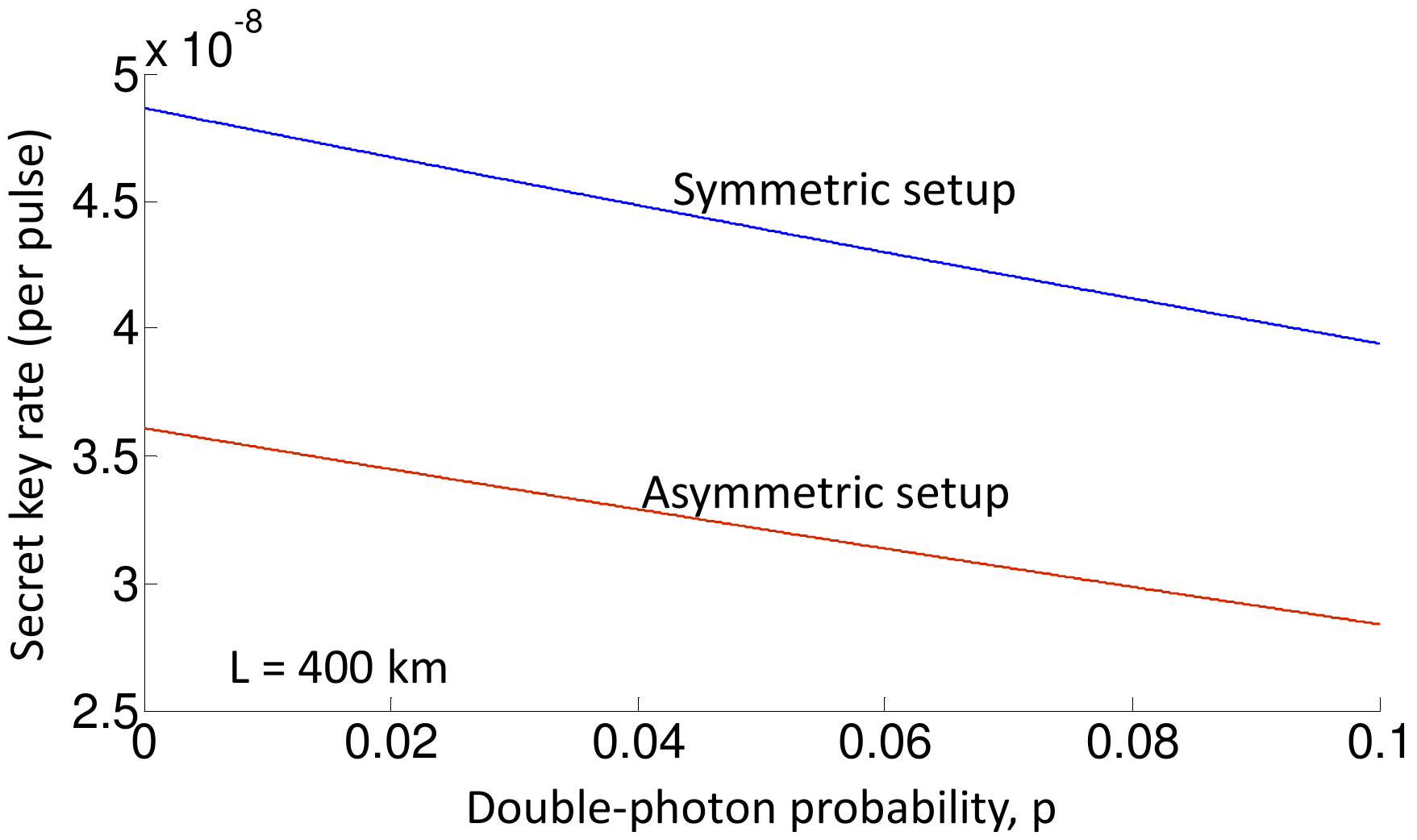}
\par\end{centering}
\caption{\label{fig:versus-p_q} Secret key generation rate per transmitted pulse versus the double-photon
probability, $p$. In all curves $L=$ 400~km and all other parameters are taken from Table~\ref{tab:Nominal-values-used}. In the symmetric case, $L_A = L_B$, whereas in the asymmetric case, $L_A = L$ and $L_B =0$. }
\end{figure}

Figure \ref{fig:versus-p_q} shows the secret key generation rate per transmitted pulse for the setup of Fig.~\ref{fig:Diagram-for-MDI-QKD} versus the double-photon probability. We have used a nominal set of values, listed in Table~\ref{tab:Nominal-values-used}, for all relevant parameters. The near-ideal nominal values for
quantum efficiency and dark count have been achieved in \cite{Nam_NatPhot_93p_2013}. We have considered two scenarios. The first is a symmetric setup, when the BSM module is located in the middle of the link, i.e., $L_A = L_B$. The other scenario is for when the BSM module is next to the Bob's apparatus, similar to the situation that we have in the side-BSM of Fig.~\ref{Fig:setups}(b). In both cases, there seems to be little effect on the key rate as a result of introducing double-photons. The key reason for this behavior is the fact that the only error term in \eqref{eq:Rss} that depends on $p$ is $E_{pp}^z$. An error in the $z$ basis arises from the cases where Alice and Bob are both sending the same bits, let's say both send a signal in their respective $r$ modes, but one $r$ detector and one $s$ detector clicks in Fig.~\ref{fig:Diagram-for-MDI-QKD}. The click on the $s$ detectors should then be because of dark counts and is not affected by the double photon states in the $r$ modes. Double photons slightly change the rate, as we disregard double-click cases, and that is the reason for lower key rates once $p$ increases. 

\section{MDI-QKD with ensemble-based memories}

In this section, we analyze the effect of multiple excitations in \eqref{init_joint} on the key rate of the memory-assisted MDI-QKD link of Fig.~\ref{Fig:setups}(b). We again use the phase-encoding scheme described in Sec.~\ref{Sec:PhaseEnc} and combine it with four ensemble-based memories as described below. In contrast to the previous section, where double-photon terms had little effect on system performance, it turns out that, within the setup of Fig.~\ref{Fig:setups}(b), multiple excitations in memories would adversely affect the achievable key rate. We then look at the scheme of Fig.~\ref{Fig:setups}(c) and show, how, in principle, we can remedy this problem.

\subsection{Setup description}

\begin{figure}
\begin{centering}
\includegraphics[width=8.6cm]{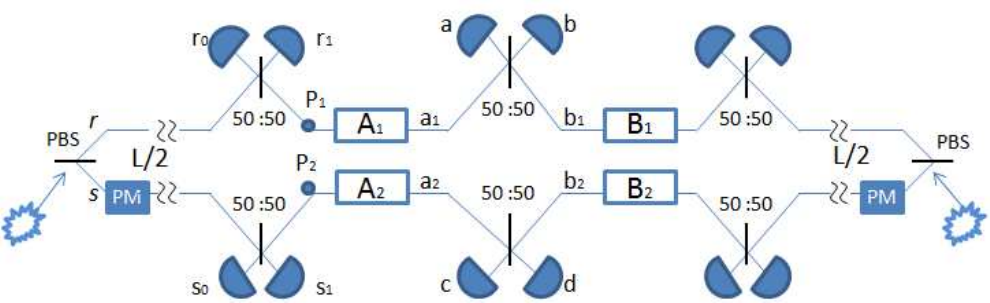}
\par\end{centering}
\caption{{\footnotesize \label{fig:new scheme}Schematic diagram for the MDI-QKD setup with ensemble-based memories, represented by $A_1$, $A_2$, $B_1$, and $B_2$.}}
\end{figure}

Figure \ref{fig:new scheme} shows the phase-encoding variant of the memory-assisted MDI-QKD system of Fig.~\ref{Fig:setups}(b). Here, in order to focus on the memory effects, we assume Alice and Bob are using perfect single-photon sources. For each photon encoded and sent by the users, we pump the corresponding memories $A_1$, $A_2$, $B_1$, and $B_2$ in order to generate a joint photonic-atomic state as in \eqref{init_joint}. The state sent by the user is indirectly loaded to the memories by the side-BSM modules in Fig.~\ref{fig:new scheme}. For instance, on the Alice side, we perform a BSM on the single-photon state sent by Alice and $P_1$ and $P_2$ states using the same BSM module as in Fig.~\ref{fig:Diagram-for-MDI-QKD}. A successful side BSM, with the same definition for success as in Sec.~\ref{Sec:PhaseEnc}, would ideally load the memory with a state corresponding to what the users have sent. For instance, if Alice uses the $z$ basis, and sends a signal in the $r$ mode, a successful BSM on her side, would imply that the memories $A_1$-$A_2$ are ideally in the $|01\rangle_{A_1 A_2}$ state. Of course, considering the dark current and double-photon terms, we will deviate from this ideal case, and that is what we are going to study in this paper. Alice and Bob attempt repeatedly to load their memories until they succeed, at which point they wait for the other party to complete this task. Once both sets of memories are loaded, we read out all four memories and proceed with the middle BSM. Once the results of all three BSMs as well as the bases used are communicated to users, Alice and Bob can  distill with a sifted key bit. Table~\ref{Tab:BitPattern} shows what bits Alice and Bob assign to their sifted keys depending on the results of the three BSM operations.

\begin{table}
\footnotesize
\begin{centering}
\vspace{2mm}
\begin{tabular}{|c|c|c|c|c|}
\hline 
Basis & Alice BSM & Bob BSM & Middle BSM & Bit assignment \\
\hline 
$z$ & type I/II & type I/II & type I/II & Bob flips his bit \\
\hline 
$x$ & type I (II) & type I (II) & type I & Bob keeps his bit \\
\hline 
$x$ & type I (II) & type I (II) & type II & Bob flips his bit \\
\hline 
$x$ & type I (II) & type II (I) & type I & Bob flips his bit \\
\hline 
$x$ & type I (II) & type II (I) & type II & Bob keeps his bit \\
\hline 
\end{tabular}
\par\end{centering}
\caption{{\footnotesize \label{Tab:BitPattern} Bit assignment protocol as a function of the results of the three BSMs in Fig.~\ref{fig:new scheme}. Here, Alice (Bob) BSM refers to the side BSM on the left (right).}}
\end{table}

\subsection{Key rate analysis}
In this section, the key rate for the setup of Fig.~\ref{fig:new scheme} is obtained under the normal operation condition when no eavesdropper is present. Using the efficient QKD protocol, where the $z$ basis is used more often than the $x$ basis, the secret key rate per transmitted pulse is lower bounded by
\begin{equation}
\label{Rate:QM}
R_{\rm QM}\geq Y_{11}^{\rm QM}\left[1- h\left(e_{11;x}^{\rm QM}\right) - h \left(e_{11;z}^{\rm QM}\right)\right] ,
\end{equation}
where $e_{11;x}^{\rm QM}$ and $e_{11;z}^{\rm QM}$, respectively, represent the QBER between Alice and Bob in the $x$ and $z$ basis, when single photons are sent, and $Y_{11}^{\rm QM}$ represents the probability that, in the $z$ basis, both sets of memories $A$ and $B$ are loaded {\em and} the middle BSM is successful. In Appendix \ref{App:QM}, we derive all above terms assuming that memories may undergo amplitude decay according to an exponential law. That is, if the recall/reading efficiency, right after a successful writing procedure, is denoted by $\eta_{r0}$, the reading efficiency after a time $t$ is given by $\eta_r(t) = \eta_{r0}\exp(-t/T_1)$, where $T_1$ is the amplitude decay time constant.

In the absence of dark counts, memory decay, and source imperfections, the major source of noise in the setup of Fig.~\ref{fig:new scheme} is the multiple-excitation terms in \eqref{init_joint}. Even if the users send exactly one photon, the state loaded to the QMs may contain more than one excitation overall. These additional excited atoms will cause errors in the middle BSM setup. The errors in the latter stage are partly similar to what we studied in the previous section, when we considered imperfect single-photon sources. These cases correspond to loading states like $|20\rangle_{A_1A_2}$ into $A_1$-$A_2$ memories, or similar states for $B_1$-$B_2$. There are, however, other terms that must be considered, such as $|11\rangle_{A_1A_2}$, and they turn out to give a much larger contribution to the noise terms in \eqref{Rate:QM}. Our analysis in this section, considers up to two excitations in each memory module.  

\begin{figure}
\centering
\includegraphics[width=7.6cm]{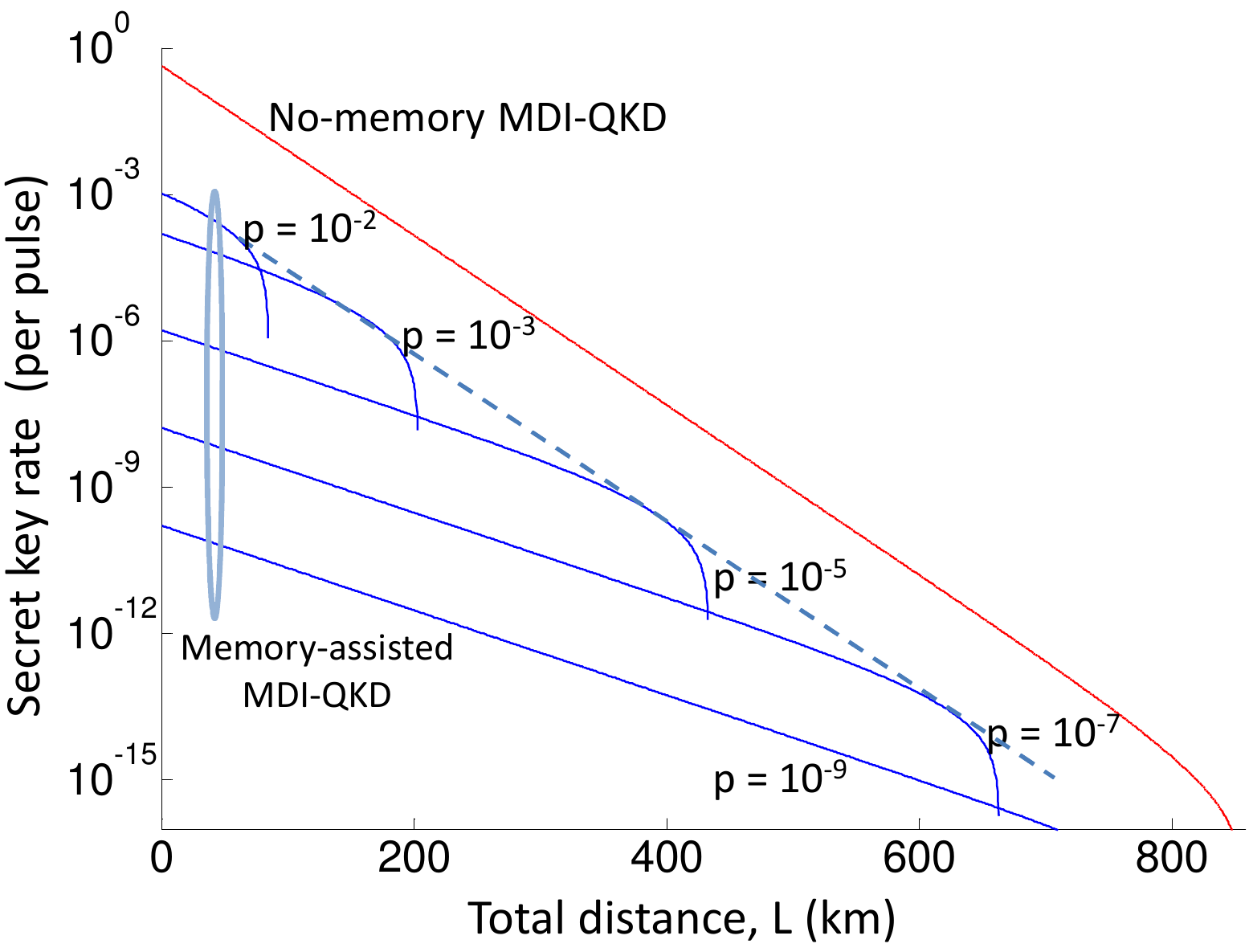}
\caption{{\footnotesize \label{fig:comparison_with_mem}Secret key generation rate per transmitted pulse versus distance for the MDI-QKD scheme in Fig.~\ref{fig:new scheme} with QMs (blue curves) and that of Fig.~\ref{fig:Diagram-for-MDI-QKD} without QMs (red curves) for different values of the excitation probability $p$. Nominal values are used as in Table~\ref{tab:Nominal-values-used} with $T_{1}=\infty$. For the no-memory curve, $L_A = L_B$ and $p=0$.}}
\end{figure}

Figure \ref{fig:comparison_with_mem} shows the effect of multiple excitations in the scheme of Fig.~\ref{fig:new scheme} and compares it with a symmetric no-memory setup as in Fig.~\ref{fig:Diagram-for-MDI-QKD}. Assuming no decay or misalignment in the setup and with a negligible amount of dark count as in Table~\ref{tab:Nominal-values-used}, Fig.~\ref{fig:comparison_with_mem} shows that the memory-assisted system of Fig.~\ref{fig:new scheme} cannot outperform the no-memory system within a reasonable range of rates and/or distances. Here, we have considered different values of $p$. As we decrease the value of $p$, the chance of entangling a photon with the memories becomes lower, and that is why the initial key generation rate drops. However, lower values of $p$ will make the generation of multiple-excitation states less likely and that is why the cut-off security distance becomes longer. The rate, however, remains below the no-QM curve even for very small values of $p$. 

In order to understand the above behavior, we need to look more closely at the dynamics of different terms in \eqref{Rate:QM}. The term $Y_{11}^{\rm QM}$ is proportional to the loading probability, i.e., the success probability in each of the side BSMs of Fig.~\ref{fig:new scheme}. In order to have a successful BSM we need to get two clicks, one on the upper arm, and one in the lower one. For short distances, the two clicks are typically caused by the photon sent by the user and a photon entangled with the two memories on each side. The loading probability, in this limit, is then on the order of $p \exp[-(L/2)/L_{\rm att}]$, where $p$ is the probability that one of the two ensembles on each side has one excitation, and $\exp[-(L/2)/L_{\rm att}]$ is the channel efficiency for the transmitted photon by the user. The initial slope of the curves in Fig.~\ref{fig:comparison_with_mem} corresponds to the above scaling with distance, similar to that of quantum repeaters. As the distance becomes longer and longer, the chance of receiving the photon sent by the user becomes slimmer and slimmer. In this limit, a successful BSM is often caused by photons originating from memories, in particular, terms like $|11\rangle_{A_1A_2}|11\rangle_{P_1P_2}$. Such successful BSMs do not imply any correlations between the states of memories and that of Alice or Bob, and will simply result in random errors and the eventual decline of the key rate to zero. Given that the probability of generating a two-photon state is on the order of $p^2$, the transition from the first region to the cut-off region roughly occurs at a distance $L_c$, where $p \exp[-(L_c/2)/L_{\rm att}] \approx p^2$, or equivalently, when $\exp[-(L_c/2)/L_{\rm att}] \approx p$.
This implies that the total rate would then scale as $p \exp[-(L_c/2)/L_{\rm att}] \approx \exp[-L_c/L_{\rm att}]$, which is similar to a no-QM system. This is evident in Fig.~\ref{fig:comparison_with_mem} by the envelop (dashed line) of QM-assisted curves, which is parallel to the no-QM curve. Considering the additional inefficiencies in the memory-assisted system as compared to the no-QM one, for the range of values used in our calculations, it becomes practically impossible to beat the no-QM system if we use ensemble-based memories in the setup of Fig.~\ref{fig:new scheme}. Note that the performance would further degrade if memory decay effects are also included.



\subsection{Modified Setup}
The results of the previous subsection imply that ensemble-based memories barely offer any advantages over no-QM systems within the setup of Fig.~\ref{fig:new scheme}. The key reason is the generation of multiple excitation terms in the memory once a photonic state is entangled with it via driving off-resonant Raman transitions. In the scheme of Fig.~\ref{Fig:setups}(b), we use the entanglement between the memory and the photon to effectively {\em teleport}, via the side BSM, the user's state onto the memories. This task can be done in a different way as shown in Fig.~\ref{Fig:setups}(c). In this setup, we use an EPR source to generate entangled photons. If we store one of the photons into the memory, we would have effectively achieved the same required entanglement between the memory and the other photon in the EPR pair, and the rest of the protocol can proceed as before. Note that this scheme is not fully heralding, because we cannot tell if the photon has actually been stored in the QM, but considering that entangled photons are generated locally, the required writing procedure can be very efficient \cite{HighStorageEff_PRL13}. 

The main advantage that the setup of Fig.~\ref{Fig:setups}(c) offers is its in-principle resilience to multi-photon terms. If the employed EPR sources do not include multi-photon terms, we only generate at most one excited atom in the respective ensembles. That implies that once we read the memories, there will only be one photon from each side and we will not deal with the types of errors that exist in the setup of Fig.~\ref{fig:new scheme}.

Another advantage of the setup of Fig.~\ref{Fig:setups}(c) is that we are not, in this setup, restricted by the writing time of the memories. The writing time specifies the repetition rate for the setup of Figs.~\ref{Fig:setups}(b) and \ref{fig:new scheme}. If we need to repeatedly write into a memory, the writing time will be restricted by the time it takes for possible cooling operations or when we need to pump the QM to a special initial state. This will in essence reduce the key generation rate per unit of time. In Fig.~\ref{Fig:setups}(c), we can avoid sequential writing into the QMs if we use a delay line and a fast optical switch for the photon that must be stored into the memory. We will only attempt to write into the memory once there is a successful side BSM. In this way, the overhead time for preparing the memory will become almost irrelevant, and the repetition rate is determined by the EPR source entanglement generation rate. Note that the delay time required in the above scheme is typically much shorter than the required storage time in the memory. One can, however, study the system performance when an optical memory (delay line), rather than a QM, is in use. Alternatively, one may drive a large number of these fully optical systems to asymptotically get the same rate improvement as obtained here \cite{Azuma_adaptiveBSM}. In either case, a single memory-assisted system is expected to outperform a single all-optical system over a certain range.

 The choice of the EPR source is very important in the scheme of Fig~\ref{Fig:setups}(c). In particular, it is important to note that the existing sources of entangled photons based on parametric down-conversion are not suitable for this scheme. In fact, they have exactly the same multi-photon statistics as given by \eqref{init_joint} for the number of photons in their idler and signal beams \cite{Razavi_CohMeas_PRA09}, hence would give the same kind of performance as in Fig.~\ref{fig:comparison_with_mem}. Quantum-dot based sources, on the other hand, offer high generation rates of entangled states with negligible two-photon components \cite{QDot_entg_low_g2:NatPhot2014, QDot_entg_with_cavity:Nat2010}. They need, nevertheless, to improve their fidelity of generated entangled photons \cite{QDotToshibaPRL2007}. The performance of MDI-QKD systems relying on such imperfect sources will be investigated in a separate publication.

\begin{figure}[htbp]
  \centering
  \includegraphics[width=7.5cm]{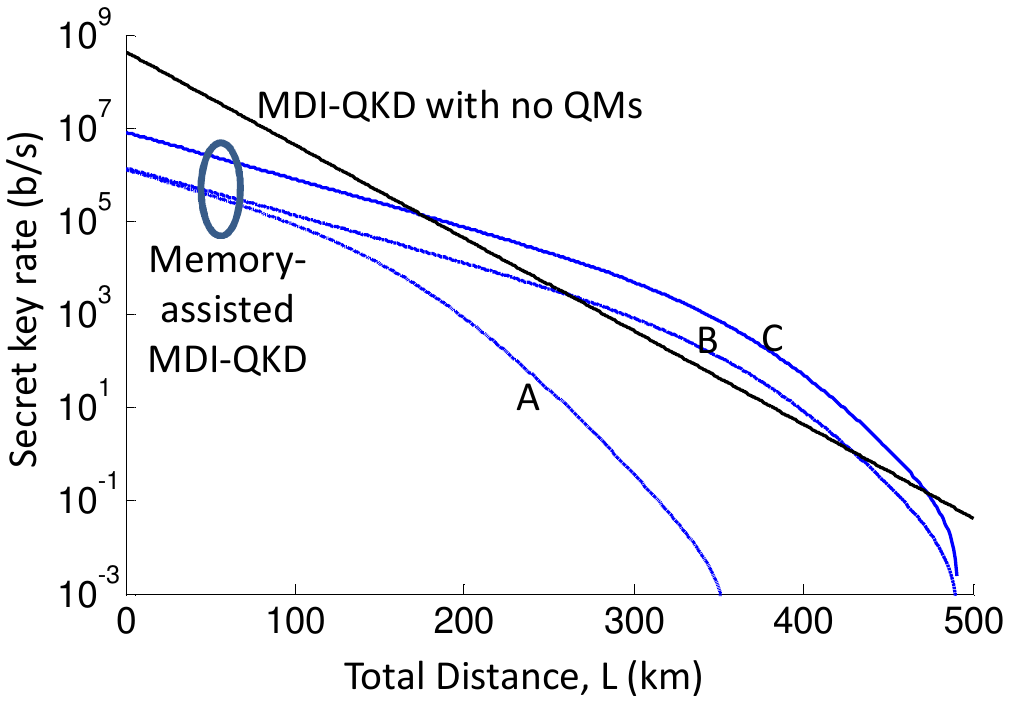}
\caption{\label{fig:rate_mod}Secret key generation rate for the scheme in Fig.~\ref{Fig:setups}(c) using ensemble-based QMs. Ideal EPR sources with 12\% efficiency are used. Curve A assumes $T_1 = T_2 = 1.5$ $\mu$s, where $T_2$ is the dephasing time constant, and the initial retrieval efficiency is $\eta_{r0}=0.3$; curve B assumes $T_1 = T_2 = 150$ $\mu$s and $\eta_{r0}=0.3$; and curve C assumes $T_1 = T_2 = $ 150~$\mu$s and $\eta_{r0}=0.73$. In all curves, reading and writing times are 300 ps, the repetition rate is 1 GHz, channel loss is 0.2~dB/km, and detector parameters are as in Table I.}
\end{figure}

In this section, we use the results reported in \cite{Panayi_NJP2014} to find the key rate for the setup of Fig.~\ref{Fig:setups}(c) assuming that the EPR source generates a maximally entangled state. Figure~\ref{fig:rate_mod} shows the achievable key rates for the scheme of Fig.~\ref{Fig:setups}(c), when it is driven by an EPR source with 12\% efficiency \cite{QDot_entg_with_cavity:Nat2010}. We have neglected the double-photon emissions and have assumed that each generated photon can be loaded into the memory with unity efficiency. In Fig.~\ref{fig:rate_mod}, curve A is based on realistic parameter values as reported in \cite{Walmsley:PRL:2010}. The achievable key rate can clearly not beat the no-QM system. By improving the coherence time of the QMs by two orders of magnitude, as in curve B, we can now outperform the no-memory system over a certain range. This range becomes wider and more practical, as shown in curve C, if our initial retrieval efficiency is increased from 0.3 to 0.73. Both required improvements in curve C are potentially achievable within our current technology as they have been obtained in other similar setups \cite{Pan:NatPhys:2012} for cold atomic ensembles. This promises an imminent exploitation of QMs in real systems with clear advantages over no-memory systems.


\section{Conclusion}
In this paper, we provided a full analysis of the MDI-QKD systems that use ensemble-based memories. Memory-assisted MDI-QKD is expected to beat conventional no-memory QKD links in rate and distance. This is to be achieved without requiring much demanding technology for quantum memories, which hinders the progress of quantum repeaters. In memory-assisted MDI-QKD, memories are required to be fast and to demonstrate sufficiently long coherence times as compared to their access times. Both these conditions have been met for certain memories that rely on atomic ensembles or atomic frequency combs. In both cases, the memories, when driven by coherent pulses, suffer from multiple excitation effects. In this paper, we showed that these multiple excitations deteriorate the performance of certain memory-assisted MDI-QKD systems to the extent that they could no longer beat their no-memory counterparts. We showed that in order to revive the promised advantage of beating no-memory systems, using ensemble-based memories, one needed to be equipped with almost ideal entangled-photon sources. In other words, our memory problem would be converted into a source problem. The prospect of developing memory-assisted QKD systems is, nevertheless, still bright. In particular, sources based on quantum dot structures have shown to have very little multi-photon components, and can be run at GHz rates. Further progress in that ground put together with the slight improvements that we need on the memory side would enable us to devise the first generation of memory-assisted systems that offer realistic advantages in practice.


%

\appendix
\section{MDI-QKD with imperfect sources}
\label{App:no-QM}
In this Appendix we will derive the terms in \eqref{eq:Rss} for the setup of Fig.~\ref{fig:Diagram-for-MDI-QKD}, considering path loss, quantum efficiency $\eta_{d},$ dark count rates $d_{c}$, double-photon probability $p$, and misalignment probability $e_d$ assuming that no eavesdropper is present. This provides us with an estimate of how well the system performs under normal conditions. In \eqref{eq:Rss}, $Y_{11}^{z}$ and $e_{11}^{x}$ have already been calculated in \cite{MXF:MIQKD:2012}. Here, we will derive the other two terms $Q_{pp}^z$ and $E_{pp}^z$. In the $z$ basis, a successful click event at the BSM module corresponds to different key bits at Alice's and Bob's ends. We can therefore separate the input states that result in correct inference of bits versus those causing errors. The input states that result in correct inference of bits are those that correspond to sending different bits by Alice and Bob given by
\begin{equation}
\rho_{C}^{(\rm in)}= [\rho_{r_A}(p) \otimes \rho_{s_B}(p) + \rho_{s_A}(p) \otimes \rho_{r_B}(p)]/2,
\label{eq:initial_dens_mat_1}
\end{equation}
whereas
\begin{equation}
\rho_{E}^{(\rm in)}= [\rho_{r_A}(p) \otimes \rho_{r_B}(p) + \rho_{s_A}(p) \otimes \rho_{s_B}(p)]/2
\label{eq:initial_dens_mat_2}
\end{equation}
results in erroneous decisions. In above equations, $r_{A(B)}$ and $s_{A(B)}$ subscripts, respectively, refer to the $r$ and $s$ optical modes of Alice (Bob) in Fig.~\ref{fig:Diagram-for-MDI-QKD}. Note that terms corresponding to $O\left(p^{2}\right)$ are neglected in \eqref{eq:initial_dens_mat_1} and \eqref{eq:initial_dens_mat_2}. Each of the above states undergoes a state transformation according to the butterfly module in Fig.~\ref{fig:BSM}(b). We denote this transformation by $B_{\eta_a,\eta_b}^{xy}$, where $x$ and $y$ refer to the input modes to the module. The input-output relationships for this butterfly operation are given in Table~\ref{Tab:B05eta} for a range of input states of interest. The output states in Fig.~\ref{fig:BSM}(b), for the input states as in \eqref{eq:initial_dens_mat_1} and \eqref{eq:initial_dens_mat_2}, are then given by
\begin{equation}
\rho_{K}^{(\rm out)}=B_{\eta_a,\eta_{b}}^{r_A r_B} \otimes B_{\eta_a,\eta_{b}}^{s_A s_B} (\rho_{K}^{(\rm in)}), \quad\mbox{$K=C,E$,}
\label{eq:butterfly}
\end{equation}
where $\eta_{a}=\eta_{\rm ch}(L_A) \eta_{d}$ and $\eta_{b}=\eta_{\rm ch}(L_B) \eta_{d}$.

\begin{table*}

\begin{footnotesize} %
\begin{centering}
\vspace{2mm}
\begin{tabular}{|c|c|}
\hline 
$\rho_{AB}$  & $B_{\eta_{a},\eta_{b}}^{AB}\left(\rho_{AB}\right)$\tabularnewline
\hline 
$|10\left\rangle \left\langle 10|\right.\right.$  & $\frac{\eta_{a}}{2}\left(|10\left\rangle \left\langle 10|\right.\right.+|01\left\rangle \left\langle 01|\right.\right.\right)+\left(1-\eta_{a}\right)|00\left\rangle \right\langle 00|$\tabularnewline
\hline 
$|01\left\rangle \left\langle 01|\right.\right.$  & $\frac{\eta_{b}}{2}\left(|10\left\rangle \left\langle 10|\right.\right.+|01\left\rangle \left\langle 01|\right.\right.\right)+\left(1-\eta_{b}\right)|00\left\rangle \right\langle 00|$\tabularnewline
\hline 
$|11\left\rangle \left\langle 11|\right.\right.$  & $\frac{1}{2}\left(\eta_{a}+\eta_{b}-2\eta_{a}\eta_{b}\right)\left(|10\left\rangle \left\langle 10|\right.\right.+|01\left\rangle \left\langle 01|\right.\right.\right)+\left(1-\eta_{a}\right)\left(1-\eta_{b}\right)|00\left\rangle \right\langle 00|+\frac{\eta_{a}\eta_{b}}{2}\left(|20\left\rangle \left\langle 20|\right.\right.+|02\left\rangle \left\langle 02|\right.\right.\right)$\tabularnewline
\hline 
$|20\left\rangle \left\langle 20|\right.\right.$  & $\eta_{a}\left(1-\eta_{a}\right)\left(|10\left\rangle \left\langle 10|\right.\right.+|01\left\rangle \left\langle 01|\right.\right.\right)+\left(1-\eta_{a}\right)^{2}|00\left\rangle \right\langle 00|+\frac{\eta_{a}^{2}}{4}\left(|20\left\rangle \left\langle 20|\right.\right.+|02\left\rangle \left\langle 02|\right.\right.\right)$\tabularnewline
\hline 
$|02\left\rangle \left\langle 02|\right.\right.$  & $\eta_{b}\left(1-\eta_{b}\right)\left(|10\left\rangle \left\langle 10|\right.\right.+|01\left\rangle \left\langle 01|\right.\right.\right)+\left(1-\eta_{b}\right)^{2}|00\left\rangle \right\langle 00|+\frac{\eta_{b}^{2}}{4}\left(|20\left\rangle \left\langle 20|\right.\right.+|02\left\rangle \left\langle 02|\right.\right.\right)$\tabularnewline
\hline 
$|21\left\rangle \left\langle 21|\right.\right.$  & $\eta_{c}\left(1-\eta_{a}\right)\left[\eta_{a}\left(1-\eta_{b}\right)+\frac{\eta_{b}}{2}\left(1-\eta_{a}\right)\right]\left(|10\left\rangle \left\langle 10|\right.\right.+|01\left\rangle \left\langle 01|\right.\right.\right)+\left(1-\eta_{a}\right)^{2}\left(1-\eta_{b}\right)|00\left\rangle \right\langle 00|$\tabularnewline
 & $+\eta_{a}\left[\frac{\eta_{a}}{4}\left(1-\eta_{b}\right)+\eta_{b}\left(1-\eta_{a}\right)\right]\left(|20\left\rangle \left\langle 20|\right.\right.+|02\left\rangle \left\langle 02|\right.\right.\right)+\frac{3}{8}\eta_{a}^{2}\eta_{b}\left(|30\left\rangle \left\langle 30|\right.\right.+|03\left\rangle \left\langle 03|\right.\right.\right)$\tabularnewline
\hline 
$|12\left\rangle \left\langle 12|\right.\right.$  & $\left(1-\eta_{b}\right)\left[\eta_{b}\left(1-\eta_{a}\right)+\frac{\eta_{a}}{2}\left(1-\eta_{b}\right)\right]\left(|10\left\rangle \left\langle 10|\right.\right.+|01\left\rangle \left\langle 01|\right.\right.\right)+\left(1-\eta_{b}\right)^{2}\left(1-\eta_{a}\right)|00\left\rangle \right\langle 00|$\tabularnewline
 & $+\eta_{b}\left[\frac{\eta_{b}}{4}\left(1-\eta_{a}\right)+\eta_{a}\left(1-\eta_{b}\right)\right]\left(|20\left\rangle \left\langle 20|\right.\right.+|02\left\rangle \left\langle 02|\right.\right.\right)+\frac{3}{8}\eta_{a}\eta_{b}^{2}\left(|30\left\rangle \left\langle 30|\right.\right.+|03\left\rangle \left\langle 03|\right.\right.\right)$\tabularnewline
\hline 
$|10\left\rangle \left\langle 01|\right.\right.$  & $\frac{1}{2}\sqrt{\eta_{a}\eta_{b}}\left(|10\left\rangle \left\langle 10|\right.\right.-|01\left\rangle \left\langle 01|\right.\right.\right)$\tabularnewline
\hline 
$|01\left\rangle \left\langle 10|\right.\right.$  & $\frac{1}{2}\sqrt{\eta_{a}\eta_{b}}\left(|10\left\rangle \left\langle 10|\right.\right.-|01\left\rangle \left\langle 01|\right.\right.\right)$\tabularnewline
\hline 
$|11\left\rangle \left\langle 20|\right.\right.$  & $\left(1-\eta_{a}\right)\sqrt{\frac{\eta_{a}\eta_{b}}{2}}\left(|10\left\rangle \left\langle 10|\right.\right.-|01\left\rangle \left\langle 01|\right.\right.\right)+\frac{\eta_{a}\sqrt{\eta_{a}\eta_{b}}}{2\sqrt{2}}\left(|20\left\rangle \left\langle 20|\right.\right.-|02\left\rangle \left\langle 02|\right.\right.\right)$\tabularnewline
\hline 
$|11\left\rangle \left\langle 02|\right.\right.$  & $\left(1-\eta_{a}\eta_{c}\right)\sqrt{\frac{\eta_{a}\eta_{b}}{2}}\left(|10\left\rangle \left\langle 10|\right.\right.-|01\left\rangle \left\langle 01|\right.\right.\right)+\frac{\eta_{a}\sqrt{\eta_{a}\eta_{b}}}{2\sqrt{2}}\left(|20\left\rangle \left\langle 20|\right.\right.-|02\left\rangle \left\langle 02|\right.\right.\right)$\tabularnewline
\hline 
$|20\left\rangle \left\langle 11|\right.\right.$  & $\left(1-\eta_{a}\right)\sqrt{\frac{\eta_{a}\eta_{b}}{2}}\left(|10\left\rangle \left\langle 10|\right.\right.-|01\left\rangle \left\langle 01|\right.\right.\right)+\frac{\eta_{a}\sqrt{\eta_{a}\eta_{b}}}{2\sqrt{2}}\left(|20\left\rangle \left\langle 20|\right.\right.-|02\left\rangle \left\langle 02|\right.\right.\right)$\tabularnewline
\hline 
$|02\left\rangle \left\langle 11|\right.\right.$  & $\left(1-\eta_{a}\right)\sqrt{\frac{\eta_{a}\eta_{b}}{2}}\left(|10\left\rangle \left\langle 10|\right.\right.-|01\left\rangle \left\langle 01|\right.\right.\right)+\frac{\eta_{a}\sqrt{\eta_{a}\eta_{b}}}{2\sqrt{2}}\left(|20\left\rangle \left\langle 20|\right.\right.-|02\left\rangle \left\langle 02|\right.\right.\right)$\tabularnewline
\hline 
$|20\left\rangle \left\langle 02|\right.\right.$  & $\frac{\eta_{a}\eta_{b}}{4}\left(|20\left\rangle \left\langle 20|\right.\right.+|02\left\rangle \left\langle 02|\right.\right.\right)$\tabularnewline
\hline 
$|02\left\rangle \left\langle 20|\right.\right.$  & $\frac{\eta_{a}\eta_{b}}{4}\left(|20\left\rangle \left\langle 20|\right.\right.+|02\left\rangle \left\langle 02|\right.\right.\right)$\tabularnewline
\hline 
 & $\left(1-\eta_{a}\right)^{2}\left(1-\eta_{b}\right)^{2}|00\left\rangle \right\langle 00|+$\tabularnewline
 & $\left(1-\eta_{a}\right)\left(1-\eta_{b}\right)\left[\eta_{a}\left(1-\eta_{b}\right)+\eta_{b}\left(1-\eta_{a}\right)\right]\left(|10\left\rangle \left\langle 10|\right.\right.+|01\left\rangle \left\langle 01|\right.\right.\right)+$\tabularnewline
$|22\left\rangle \left\langle 22|\right.\right.$  & $\frac{3}{4}\eta_{a}\eta_{b}\left[\eta_{a}\left(1-\eta_{b}\right)+\eta_{b}\left(1-\eta_{a}\right)\right]\left(|30\left\rangle \left\langle 30|\right.\right.+|03\left\rangle \left\langle 03|\right.\right.\right)+$\tabularnewline
 & $\frac{1}{4}\left[\eta_{a}^{2}\left(1-\eta_{b}\right)^{2}+\eta_{b}^{2}\left(1-\eta_{a}\right)^{2}\right]\left(|20\left\rangle \left\langle 20|\right.\right.+|02\left\rangle \left\langle 02|\right.\right.\right)+\frac{3}{8}\eta_{a}^{2}\eta_{b}^{2}\left(|40\left\rangle \left\langle 40|\right.\right.+|04\left\rangle \left\langle 04|\right.\right.\right)|$\tabularnewline
\hline 
\end{tabular}

 \caption{{\label{Tab:B05eta}The input-output relationship for the asymmetric butterfly module of Fig.~\ref{fig:BSM}(b). For the sake of brevity, here,
we have only included the terms that provide us with nonzero values
after applying the measurement operation. More specifically, we have
removed all {\em asymmetric} density matrix terms, such as $|10\left\rangle \left\langle 01|\right.\right.$
or $|01\left\rangle \left\langle 10|\right.\right.$, for which the
bra state is different from the ket state, from the output state.
}}
\par\end{centering}
\end{footnotesize} 

\end{table*}
With the above output states in hand, one just needs to apply the relevant measurement operators to find all probabilities of interest. In particular, by denoting the probability that detectors $r_i$ and $s_j$, $i,j = 0,1$, click by
\begin{equation} 
P_{r_i s_j}^{(K)} = {\rm tr} (\rho_{K}^{(\rm out)} M_{r_i} M_{s_j}), \quad\mbox{$K=C,E$},
\label{eq:Prs}
\end{equation}
the probability that an acceptable click pattern occurs in the $z$
basis, $Q_{pp}^{z}$, is given by
\begin{equation}
Q_{pp}^{z} = Q^z_{C}+ Q^z_{E}
\end{equation}
where
\begin{equation}
Q^z_{K}=\left(P_{r_{0}s_{0}}^{(K)} + P_{r_{1}s_{1}}^{(K)}+P_{r_{0}s_{1}}^{(K)}+P_{r_{1}s_{0}}^{(K)}\right)/2, \quad\mbox{$K=C,E$}.
\label{eq:Yc}
\end{equation}
Finally, $E_{pp}^{z}$ is given by
\begin{equation}
E_{pp}^{z}=\frac{Q_{EE}^z}{Q_{pp}^{z}}
\end{equation}
where $Q_{EE}^z=e_{d}Q_{C}^z+(1-e_{d})Q_{E}^z.$ 

More generally, for any input state $\rho^{(\rm in)}=\rho_{r_A r_B s_A s_B}$, and for total transmissivities $\eta_A$ and $\eta_B$ for, respectively, Alice's and Bob's photons, we can define a gain parameter $Q^\beta(\eta_A,\eta_B; \rho_{r_A r_B s_A s_B})$ to represent the success probability, in basis $\beta = x,z$, for the BSM operation in Fig.~\ref{fig:Diagram-for-MDI-QKD}. For any such input state, the probabilities of getting a click on detectors $r_i$ and $s_j$, $i,j = 0,1$, is given by
\begin{equation} 
P_{r_i s_j}(\rho^{(\rm in)}) = {\rm tr} (\rho^{(\rm out)} M_{r_i} M_{s_j}),
\label{eq:PrsGen}
\end{equation}
where 
\begin{equation}
\rho^{(\rm out)}=B_{\eta_A,\eta_{B}}^{r_A r_B} \otimes B_{\eta_A,\eta_{B}}^{s_A s_B} (\rho^{(\rm in)}).
\label{eq:butterflyGen}
\end{equation}
With the above notation, we obtain
\begin{equation}
\begin{array}{c}
Q^{\beta}(\eta_{A},\eta_{B};\rho^{({\rm in)}})=P_{r_{0}s_{0}}(\rho^{({\rm in)}})+P_{r_{1}s_{1}}(\rho^{({\rm in)}})\\
+P_{r_{0}s_{1}}(\rho^{({\rm in)}})+P_{r_{1}s_{0}}(\rho^{({\rm in)}}).
\end{array}
\label{eq:Qb}
\end{equation}
The total gain for the basis $\beta = x,z$ is then given by
\begin{equation}
\label{Qbeta}
Q^{\beta}\left(\eta_{A},\eta_{B}\right)=\sum_{\mathrm{all\, input\, states\,\rho}}Q^{\beta}\left(\eta_{A},\eta_{B};\rho\right)\mathrm{Pr\left(\rho\right)}\end{equation}
Similarly, we also define $Q_C^\beta(\eta_A,\eta_B)$ to be the probability to get a successful BSM {\em and} Alice and Bob end up with correct inference of their bits: 
\begin{equation}
Q_C^\beta(\eta_A,\eta_B) = \sum_{\mbox{\footnotesize all input states $\rho$}}\negthickspace\sum_{\footnotesize \stackrel{\mbox{all correct detection}}{\mbox{pairs $(r_i,s_j)$ for input $\rho$} }}\negthickspace \negthickspace \negthickspace \negthickspace P_{r_i s_j}(\rho){\rm Pr}(\rho) .
\label{QCbeta}
\end{equation}
Likewise, $Q_E^\beta(\eta_A,\eta_B) = Q^\beta(\eta_A,\eta_B) - Q_C^\beta(\eta_A,\eta_B)$ denotes the probability to get a successful BSM and Alice and Bob end up with incorrect inference of their bits. Finally, error terms can be defined as $e^\beta Q^\beta = Q_E^\beta$ calculated at the point $(\eta_A,\eta_B)$. We use the above relationships in the next Appendix.

\section{MDI-QKD with imperfect memories}
\label{App:QM}
In this Appendix we will derive the terms in \eqref{Rate:QM} for the setup of Fig.~\ref{fig:new scheme}, considering path loss, quantum efficiency $\eta_{d},$ dark count rates $d_{c}$, excitation probability $p$ of the memories, and memories' amplitude decay assuming that no eavesdropper is present. We will follow the same procedure as in Appendix \ref{App:no-QM} to separate the terms that result in error versus correct key bits. The general idea is to find the post-measurement density matrix of memories for any relevant input state upon a successful side-BSM event. Once both sets of memories are loaded, we apply the middle BSM operation and find relevant probabilities of interest. 

The setup of Fig.~\ref{fig:new scheme} can be thought of three asymmetric MDI-QKD setups, where memories link them together. The first and second systems are those that are involved with the loading process. They include the photons entangled with memories, e.g. $P_1$ and $P_2$ on Alice side, with those sent by the users. The third one is centered around the middle BSM and the photons retrieved from the memories. Here we use the general notation introduced in \eqref{eq:PrsGen}-\eqref{QCbeta} to calculate the relevant gain and error parameters. In order to do so, we need to first find the input state for the final stage of BSM. For any input state $\rho_A^{(\rm in)}$ sent by Alice, we can find the post-measurement state $\rho_A^{(\rm pm)}(r_i,s_j;\rho_A^{(\rm in)})$ of the memories $A_1$ and $A_2$ upon a click on detectors $r_i$ and $s_j$, for $i,j = 0,1$, as follows
\begin{equation}
\label{pmstate}
\rho_A^{(\rm pm)}(r_i,s_j;\rho_A^{(\rm in)}) = \frac{{\rm tr}_{P_1,P_2,r_A,s_A} (\rho_A^{(\rm out)}M_{r_i}M_{s_j})}{{\rm tr}(\rho_A^{(\rm out)}M_{r_i}M_{s_j})},
\end{equation}
where 
\begin{equation}
\rho_{A}^{(\rm out)}=B_{\eta_a,\eta_{d}}^{r_A P_1} \otimes B_{\eta_a,\eta_{d}}^{s_A P_2} (\rho_{A}^{(\rm in)} \otimes \rho_{A_1 P_1}\rho_{A_2 P_2}), 
\end{equation}
where $\rho_{A_i P_i} = |\psi\rangle_{A_i P_i}\langle \psi|$, for $i=1,2$. Similarly, one can find the post-measurement state for $B_1$-$B_2$ memories and denote it by $\rho_B^{(\rm pm)}(r_m,s_n;\rho_B^{(\rm in)})$ once detectors $r_m$ and $s_n$, for $m,n = 0,1$, click on the side BSM of Bob. The final parameter we need from the loading stage is the loading probability, i.e., the probability to get a successful side BSM on Alice's ($K=A$) or Bob's ($K=B$) side given by
\begin{equation}
P_{K} = Q^z(\eta_{\rm ch}(L_K)\eta_d, \eta_d; |10\rangle_{r_Ks_K}\langle 10| \otimes \rho_{P_1}\rho_{P_2}) .
\end{equation}

In order to apply the middle BSM on the post-measurement states $\rho_A^{(\rm pm)}$ and $\rho_B^{(\rm pm)}$, One must consider the random nature of the loading process. Given that one set of the memories can be loaded earlier than the other, the former will undergo some amplitude decay before being read for the final BSM. That would result in an imbalanced middle BSM, where the reading efficiency for one memory could be lower than that of the other. To fully capture this random storage time, following the analysis and notations used in \cite{Panayi_NJP2014}, let us consider two geometric random 
variables $N_{A}$ and $N_{B}$ corresponding to the number of
attempts until Alice memories $\left(A_{1},\, A_{2}\right)$ and Bob
memories $\left(B_{1},\, B_{2}\right)$ are, respectively, loaded. Therefore,
the number of rounds needed to load both sets of memories will be given
by $max\left\{ N_{A},\, N_{B}\right\}$. The effective reading efficiency for memories $K=A,B$ will then be given by
\begin{equation}
\eta_{rK}=\begin{cases}
\eta_{r0}, & \mathrm{if\, memory\, K\, is\, late}\\
\eta_{r}\left(t=|N_{A}-N_{B}|T\right), & \mathrm{if\, memory\, K\, is\, early}
\end{cases}, 
\end{equation}
where $T$ is the repetition period for the protocol, determined by the writing time into memories.

With all above considerations in mind, we obtain
\begin{equation}
Y_{11}^{\rm QM}=\frac{1}{N_{L}(P_{A},P_{B})+N_r}\mathrm{E}\left\{ Q^{z}\left(\eta_{rA}\eta_d,\,\eta_{rB}\eta_d\right)\right\} 
\end{equation}
where $\mathrm{E}\left\{ \cdot\right\} $ is the expectation value
operator with respect to $N_{A}$ and $N_{B}$; $Q^{z}$ is the total gain in \eqref{Qbeta}, where the input states $\rho$ in the sum cover all possible post-measurement states that can be obtained for different states sent by Alice and Bob; and $N_{L}=\mathrm{E}\left\{ max\left(N_{A},N_{B}\right)\right\}$ and $N_r$ are obtained in \cite{Panayi_NJP2014}.

Similarly, the QBER terms in \eqref{Rate:QM} can be obtained from the following
\begin{equation}
e_{11;\beta}^{\rm QM} \mathrm{E}\left\{ Q^{\beta}\left(\eta_{rA}\eta_d,\,\eta_{rB}\eta_d\right)\right\} = \mathrm{E}\left\{ Q_E^{\beta}\left(\eta_{rA}\eta_d,\,\eta_{rB}\eta_d\right)\right\}, 
\end{equation}
where, again, the sum in \eqref{QCbeta} are taken over all possible post-measurement states obtained from \eqref{pmstate} and $\mbox{\ensuremath{\beta=x,z}}.$

Finally, to calculate the expected value terms in the above equations, one needs to use the following relationships:
\begin{eqnarray}
&S_{A<B}(\delta)=\dfrac{P_{A}P_{B}(1-P_{B})e^{-\delta}}{\left[1-(1-P_{A})(1-P_{B})\right]\left[1-(1-P_{B})e^{-\delta}\right]}& \nonumber\\
&S_{B<A}(\delta)=\dfrac{P_{A}P_{B}(1-P_{A})e^{-\delta}}{\left[1-(1-P_{A})(1-P_{B})\right]\left[1-(1-P_{A})e^{-\delta}\right]}& \nonumber\\
&\mathrm{E}\left\{ \eta_{rA}\right\} =\eta_{r0}[\dfrac{P_{B}}{1-(1-P_{A})(1-P_{B})}+S_{A<B}(T/T_{1})]& \nonumber\\
&\mathrm{E}\left\{ \eta_{rB}\right\} =\eta_{r0}[\dfrac{P_{A}}{1-(1-P_{A})(1-P_{B})}+S_{B<A}(T/T_{1})]& \nonumber
\end{eqnarray}
\begin{eqnarray}
&\mathrm{E}\left\{ \eta_{rA}\eta_{rB}\right\} =\eta_{r0}^{2}P_{0}[\dfrac{1}{1-(1-P_{A})e^{-T/T_{1}}}& \nonumber\\
&+\dfrac{1}{1-(1-P_{B})e^{-T/T_{1}}}-1]& \nonumber\\
&\mathrm{E}\left\{ \eta_{rA}^{2}\right\} =\eta_{r0}^{2}[\dfrac{P_{B}}{1-(1-P_{A})(1-P_{B})}+S_{A<B}(2T/T_{1})]& \nonumber\\
&\mathrm{E}\left\{ \eta_{rB}^{2}\right\} =\eta_{r0}[\dfrac{P_{A}}{1-(1-P_{A})(1-P_{B})}+S_{B<A}(2T/T_{1})]& \nonumber\\
&\mathrm{E}\left\{ \eta_{rA}^{2}\eta_{rB}\right\} =\eta_{r0}^{3}[P_{0}+S_{B<A}(T/T_{1})+S_{A<B}(2T/T_{1})]& \nonumber\\
&\mathrm{E}\left\{ \eta_{rA}\eta_{rB}^{2}\right\} =\eta_{r0}^{3}[P_{0}+S_{A<B}(T/T_{1})+S_{B<A}(2T/T_{1})]& \nonumber\\
&\mathrm{E}\left\{ \eta_{rA}^{2}\eta_{rB}^{2}\right\} =\eta_{r0}^{4}P_{0}[\dfrac{1}{1-(1-P_{A})e^{-T/T_{1}}}& \nonumber\\
&+\dfrac{1}{1-(1-P_{B})e^{-T/T_{1}}}-1],&
\end{eqnarray}
where $P_{A}$ ($P_{B}$) is the loading probability for Alice (Bob).


\bibliographystyle{apsrev4-1}
\bibliography{Bibli28Sept12,bib}
\end{document}